\documentclass{optica-article}
\journal{opticajournal} 

\articletype{Research Article}

\usepackage[T1]{fontenc}
\usepackage[utf8]{inputenc}

\usepackage{amsmath,amssymb}
\usepackage[ruled,vlined]{algorithm2e} 
 \usepackage{booktabs}

\begin{document}

 \title{FPGA-Based Adaptive Control for Phase Stabilization in Fiber-Optic Interferometers Using Correlated Photons}

\author{ P. M. Berto\authormark{1}, F. Campodónico\authormark{2}, A. A. Matoso\authormark{1}, S. Vergara\authormark{2,3}, P. A. Coelho\authormark{4},  G. Lima\authormark{5},  S. Pádua\authormark{1}, and J. Cariñe\authormark{2,*}}

\address{\authormark{1}Departamento de Física, Universidade Federal de Minas Gerais, Belo Horizonte, Brazil.\\
\authormark{2}Departamento de Ingeniería Eléctrica, Universidad Católica de la Santísima Concepción, Concepción, Chile.\\
\authormark{3}Centro de Energía, Universidad Católica de la Santísima Concepción, Concepción, Chile.\\
\authormark{4}Facultad de Ingeniería, Universidad San Sebastián, Concepción, 4080871, Chile.\\

\authormark{5}Departamento de Física, Universidad de Concepción, Concepción, Chile.\\
\authormark{}The authors contributed equally to this work.\\}
\email{\authormark{*}jcarine@ucsc.cl} 


\begin{abstract*} Time-bin encoded photon pairs enable robust, decoherence-resistant transmission through optical fibers for long-distance quantum communication, where phase noise poses a critical limitation to stable operation. Here, we implement an adaptive Perturbation-and-Observe algorithm on a fully digital FPGA platform operating with real-time feedback at 1 Hz. The control signal is derived from the coincidence counts of correlated photon pairs. This adaptive approach reduces the rise time by 70\% and the coincidence noise by 30\%, resulting in visibility improvements sustained for more than 600 s.These results provide an efficient solution for long-term phase stabilization in quantum and photonic systems. \end{abstract*}

 \section{Introduction}
 
The correlation between photon pairs is a non-local property inherent to quantum systems, and an effective way to generate such pairs is through Spontaneous Parametric Down-Conversion (SPDC) \cite{xavier2025energy,marcikic2002time, carvacho2015postselection,florez2015, couteau2018}. This method involves pumping a nonlinear crystal with an incident beam (pump), which, due to second-order nonlinear interactions, produces pairs of entangled photons (usually referred to as signal and idler). In this process, energy and momentum are conserved through phase matching, which can occur in two ways: (I) the signal and idler photons have linear polarization in the same direction, which is orthogonal to that of the pump beam, and (II) the pair of photons has orthogonal polarizations, with one of them matching the linear polarization of the pump \cite{revi10}. 
On the other hand, interferometry is a technique of great importance in quantum physics, enabling applications that range from studies of fundamental aspects of quantum mechanics \cite{Cosme2008, DeMartini2001, Kwiat1992, Sagioro1992, TorrezRuiz2010, Matoso2015}, such as \textit{wave-particle duality}, to the preparation and measurement of quantum states, which are of particular interest for the development of fields like quantum information\cite{Matthew2009,Gisin2002}. Among the various types of interferometers, one notable example is a modified Mach-Zehnder interferometer proposed by James D. Franson~\cite{franson1989, jin2024}, which features unbalanced arms. This design allows for testing the non-local nature of the entangled photons produced by SPDC, enabling the preparation of photonic states in a superposition of discrete temporal modes, known as time-bin states~\cite{marcikic2002time,Gisin2002}, making them highly robust against depolarization or dephasing.

However, for practical quantum applications \cite{scarani2009security}, effective phase noise control is crucial in systems such as two-arm fiber interferometers. Thermal and mechanical fluctuations induce phase drifts that reduce interference visibility and degrade overall system performance. To mitigate these effects, phase stabilization techniques have been implemented using auxiliary lasers in Mach-Zehnder and Michelson interferometers, applying techniques such as wavelength division multiplexing acting through piezoelectric or fiber stretcher devices \cite{Xavier2011,Seok2009,Grassani2014}. In these optical setups, the stabilization channel operates at a wavelength different from that of the photons of interest, which reduces control accuracy since the use of WDM channels decreases interference visibility \cite{Melo2025}, while also increasing the complexity of the system or introducing additional noise.
Perturb-and-Observe (P\&O) algorithms and PID controllers are feedback control techniques for phase stabilization in fiber optic interferometers \cite{Xavier2011, Seok2009, Grassani2014, Melo2025}. However, in practical optical systems, their performance is significantly influenced by the inherent nonlinearities and periodic nature of the interferometric response.
In such systems, phase modulation is typically achieved using elements such as fiber stretchers or piezoelectric actuators, which modulate the optical phase as a function of applied voltage. Although this relation is approximately linear over a limited operating range, it exhibits nonlinear discontinuities or phase jumps beyond certain voltage thresholds, due to saturation, hysteresis, or physical constraints of the actuator. These discontinuities result in a nonlinear and discontinuous phase-voltage relationship, which complicates the control loop design and can lead to instability or loss of phase lock when employing linear control strategies such as PID. In addition, the interferometric output signal exhibits a sinusoidal dependence on the optical phase difference between the two arms of the interferometer, described by: $I(\phi) = \frac{I_0}{2} \left[ 1 + v\cos(\Delta\phi)\right]$,
where \(\Delta\phi\) represents the optical phase difference between the arms of the interferometer and $v$ is the visibility of the interference curve. The sinusoidal nature of the response, combined with random phase shifts induced by noise or system perturbations, introduces ambiguity in the control operating point, as the signal gradient periodically changes sign. 
PID controller is a feedback-based control loop that operates on three methods: proportional to, integral of, and derivative of variables relative to the system error, becoming highly sensitive to local nonlinearities and may exhibit instability or degraded performance when operating outside the linear region of the interferometric response.
In contrast, the  P\&O does not require an explicit mathematical model of the system. Introduces incremental perturbations to the control input and observes the corresponding changes in the system output to determine the appropriate adjustment direction. This model-free approach enables robust and stable operation even in the presence of phase wrapping or nonlinear and discontinuous phase-voltage behavior.
From a hardware implementation perspective, the P\&O algorithm is particularly well suited for deployment in field-programmable gate arrays (FPGAs). Its algorithmic simplicity and low computational overhead significantly reduce logic resource utilization and timing constraints, enabling efficient, high-speed integration in embedded optical phase control systems~\cite{ricco2014fpga, Mellit2011}.
On the other hand, Abarz\'ua et al.~\cite{abarzua2024} proposed an adaptive extension of the classical Perturb-and-Observe (P\&O) algorithm for phase stabilization in multi-arm Mach–Zehnder interferometers. The algorithm dynamically adjusts the phase perturbation step size based on the observed optical response, enabling stable control without requiring an explicit mathematical model of the system, and by adapting the step size in real time, allowing the controller avoids introducing unnecessary fluctuations or noise near operating extrema and in regions close to discontinuities in the phase-voltage response, which are common in practical interferometric actuators.
In this work, we implement the adaptive P\&O  algorithm in a two-arm fiber optic interferometer designed for correlated photon pairs experiments. We compare the performance of the classical and adaptive P\&O algorithms under realistic experimental conditions and demonstrate that the adaptive version significantly improves phase stabilization, enhancing the overall precision and interference contrast of the quantum system. The algorithm was implemented in an FPGA-based control system, preserving its computational simplicity and enabling high-speed and robust feedback control suitable for quantum optical applications. 
\section{Materials and Methods}
\subsection{Optical Setup}
\begin{figure}[htbp]
\centering\includegraphics[width=13cm]{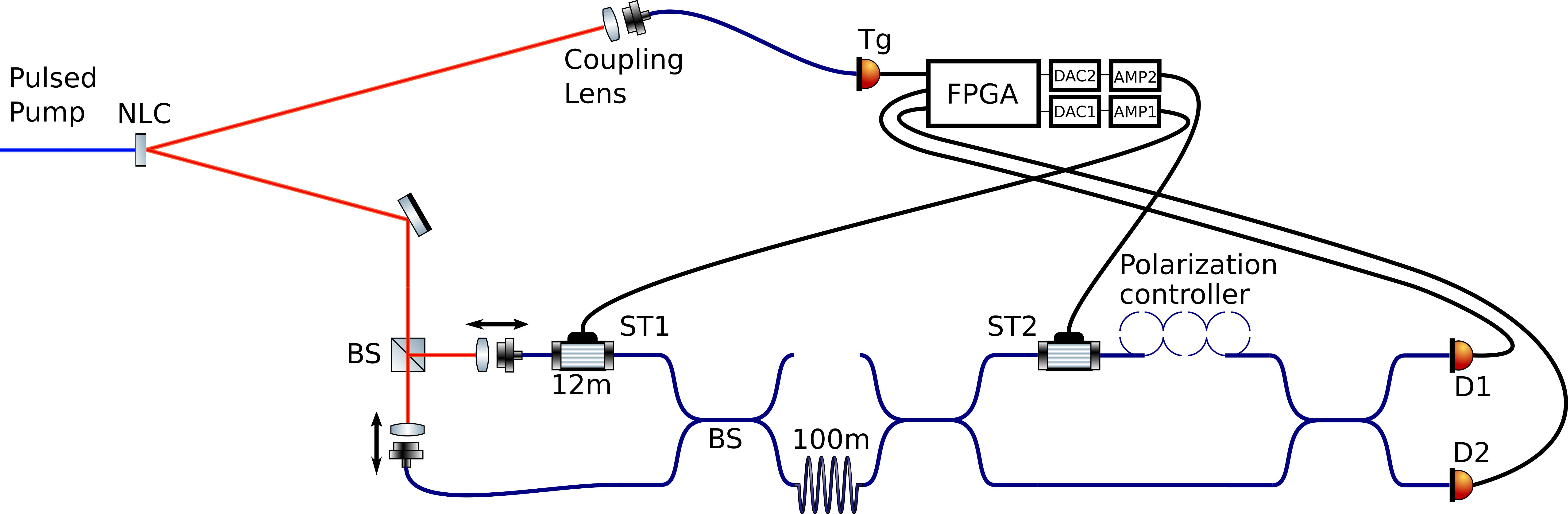}
\caption{Optical setup to generate single photons in time-bin state superpositions. Correlated photon pairs are generated by Spontaneous parametric down-conversion, being one of them detected as a trigger (Tg) in a heralded single-photon source scheme. The other goes through two Franson interferometers (unbalanced Mach-Zenhder interferometer) in sequence with fiber stretchers (ST1 and ST2) in their long arms modulating the longitudinal phase.}
\label{Fig_setup}
\end{figure}

The scheme of the experimental setup is shown in Figure \ref{Fig_setup}.

A $76$ MHz femtosecond pulsed laser at $810$ nm produces a 405 nm pulsed coherent beam by second harmonic generation that pumps a nonlinear crystal cut for type I phase match, generating two non-colinear beams of correlated photon pairs at 810 nm by SPDC. A black screen (not shown) blocks the transmitted pump beam, while photons from one of the twin photon beams are detected as a trigger by a single photon module detector (Tg). Before detection, the beam is filtered by a 10 nm bandpass interference filter (not shown in Fig. \ref{Fig_setup}) centered at 810 nm, and coupled to a monomode fiber. The other twin photon is transmitted through an identical interference filter (not shown in Fig. \ref{Fig_setup}) before it reaches the cube beam splitter. The first Franson interferometer (unbalanced Mach-Zenhder interferometer) consists of the cube beam splitter at the entrance and a fiber beam splitter at the exit, with a fiber stretcher of $\sim$ 12 m length inserted into its long arm. At one of the exit ports a $\sim$100 m fiber separates the state preparation stage performed on the first interferometer from the measurement stage performed in the second. The second Franson interferometer consists of two fiber beam splitters at the entrance/exit and another fiber stretcher into its long arm, in addition to a 3-paddle polarization controller in one arm to match the polarization when the beams recombine to interfere. Photons at each of the two exit ports are also detected by single photon module detectors D1 and D2, and their signals, along with the trigger signal, are fed to an FPGA input. The stretchers are controlled by the FPGA output depending on the single and coincidence counts processed by the FPGA, connected through two external amplifiers. With this configuration, considering the time and path degrees of freedom for the generated quantum state, one can show that the photons going through the short-long and long-short paths will exhibit complementary interference curves at D1 and D2 when detected in coincidence with the trigger properly delayed. The time delay is adjusted so that photons following the long-long and short-short paths are not detected. This interference will occur provided that the difference between the short-long/long-short paths is within the coherence length determined by the inverse of the bandwidth of the interference filters. The fiber couplers inside the first interferometer are mounted in micrometer stages that can be longitudinally displaced to match this criterion. This is the situation where we have the time-bin state superposition prepared in the first interferometer detected in the second interferometer. The function describing the behavior of coincidence counts, proportional to the fourth-order correlation function of the electric fields, is of the following form:
\begin{equation}
    C = C_0 \left[ 1 \pm v\cos(\phi_2-\phi_1+\phi_{noise})\right],
\end{equation}
where $C_0$ is the average count rate, $v$ is the visibility of the interference pattern, $\phi_{1(2)}$ is the phase added by ST1(2),  and $\phi_{noise}$ is any phase noise from uncontrolled sources such as temperature variation, mechanical vibration, and turbulence. The plus/minus sign depends on which exit port (D1 or D2) a photon is detected. This curve is the interferometric output signal mentioned above.

\subsection{Control Hardware}
\begin{figure}[htbp]
\centering\includegraphics[width=13cm]{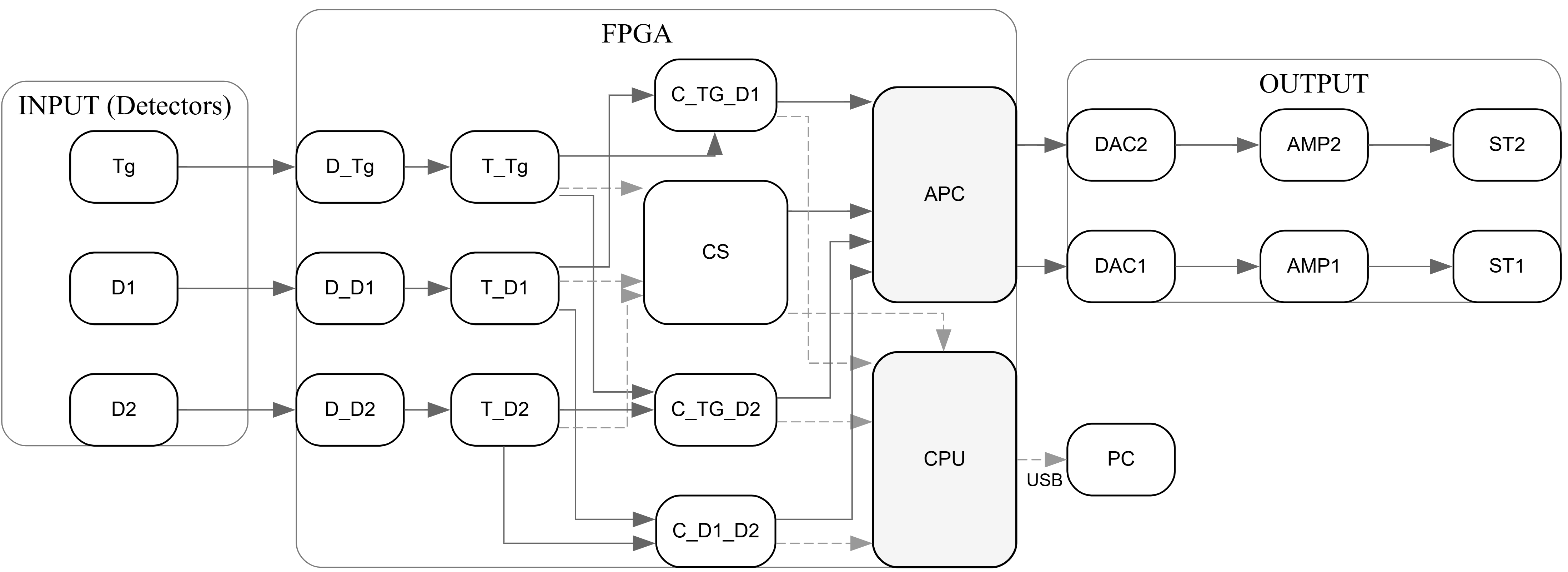}
\caption{FPGA-based phase acquisition and control. Detector signals yield singles (CS) and coincidences (CC), driving an APC module with P\&O and adaptive P\&O. Control outputs (via DAC and AMP) actuate phase stretchers (ST1/2), closing the loop. The CS and CC are also read by an embedded CPU and streamed to a host PC. Solid arrows denote feedback of control paths, and dashed arrows denote readout paths.}
\label{Fig_digramaControl}
\end{figure}

Figure \ref{Fig_digramaControl} shows the digital architecture of the system. The field programmable gate array (FPGA) interfaces with three APDs inputs: a trigger channel (Tg) that registers the timing reference (herald) and two detection channels (D1, D2) that record correlated-photon events. The FPGA phase control module generates two control outputs that are converted by digital-to-analog converters (DAC1, DAC2) operated in a range of 0V to 3V, and are externally amplified to reach more than 2$\pi$ to each ST (close to 9 V by AMP1 and AMP2), and applied to optical phase stretchers (ST1, ST2).
Each detector signal passes through a dedicated delay module ($D_Tg$, $D_D1$, $D_D2$) to compensate for propagation delays and electronic skew. The resulting aligned pulses then enter time-of-arrival (ToA) modules ($T_Tg$, $T_D1$, $T_D2$), which record each event with an internally synchronized clock and process the data as 8-bit arrival times. The effectiveness of this process was reported in \cite{cariñe2021}. Subsequently, an individual pulse counting (CS) module integrates the event rates for Tg, D1, and D2 using 24-bit counters. In parallel, three coincidence-counting (CC) modules accumulate pairwise coincidences for Tg \& D1, Tg \& D2, and D1 \& D2, each with a 24-bit counter. The CS and CC statistics are exposed both to the embedded CPU for readout and to the phase-control block for feedback.
A phase controller (APC) processes the CS and CC statistics and executes a control algorithm (P\&O or adaptive P\&O) to hold the interferometer at its operating point. The APC generates two digital control signals that drive DAC1 and DAC2, closing the feedback loop.
In parallel, the CPU packages the CS and CC records and transmits them to a host PC via USB for monitoring, logging, and higher-level processing.
\subsection{Algorithm Design}
As explained earlier, the output intensity of a two-arm interferometer can be expressed as: $I(\phi) =  \frac{I_0}{2} \left[ 1 + v\cos(\Delta\phi) \right],$ where $\Delta\phi$ is the relative phase difference between the arms and $v$ is the visibility. In a practical system, this phase difference can be written as:
\begin{equation}
\Delta\phi(t) = \phi_\text{noise}(t) - S_{ctrl}(t),
\end{equation}
with $\phi_\text{noise}(t)$ representing random phase fluctuations due to thermal, acoustic, or mechanical perturbations, and $S_{ctrl}(t)$ being the control phase applied through the actuator. The goal of the feedback controller is to adjust $S_{ctrl}(t)$ such that it cancels the noise contribution $S_{ctrl}(t) \approx \phi_\text{noise}(t)$, which drives the effective phase difference $\Delta\phi(t)=0$ toward a stable operating point, obtained $I(\phi)\approx I_0 \frac{(1+v)}{2}$, 
corresponding to constructive interference. 

Therefore, by iteratively adapting $S_{ctrl}$ to track and compensate $\phi_\text{noise}(t)$, the algorithm effectively suppresses phase noise and stabilizes the interferometer. This process ensures that the quantum interference visibility remains high, which is essential for applications such as time-bin entanglement and quantum communication protocols.

 \subsection{Algorithmic Operation}
 
To facilitate a clearer understanding, Fig.~\ref{Fig_Algoritmos} presents simplified algorithmic descriptions of the classical and adaptive Perturb-and-Observe (P\&O) control algorithms.
The classical version is illustrated in Algorithm \ref{alg_po_fpga}, where a fixed-size phase perturbation is applied and the resulting change in the interferometric output intensity is monitored. If the intensity increases, the disturbance is accepted and maintained in the same direction for the next iteration, otherwise the direction is reversed. This iterative process enables the system to converge toward a local maximum in interference contrast. 

The adaptive version extends the classical logic by dynamically adjusting the perturbation step size. Algorithm \ref{alg:adaptive_po_circular} illustrates the control strategy aimed at maximizing the system’s output intensity.

\begin{figure}[h]
\centering
{\footnotesize 
\begin{minipage}[t]{0.48\textwidth}
\begin{algorithm}[H]
\caption{ Classical Perturb-and-Observe (P\&O) implemented in FPGA. Inputs: Current intensity $I_\text{actual}$, previous intensity $I_\text{prev}$, fixed step size $p$, enable signal $EnControl$, synchronization pulse $sync$. Output: Control signal $S_{ctrl}$.}
\label{alg_po_fpga}
\SetAlgoLined
\If{rising edge of $sync$}{
    $I_\text{prev} \gets I_\text{actual}$\;
    \Switch{index}{
        \Case{0}{
            \If{$EnControl = 1$}{
                $S_{ctrl} \gets S_{ctrl} + p$\;
                index $\gets 1$\;
            }
        }
        \Case{1}{
            \If{$I_\text{actual} > I_\text{prev}$}{
                $S_{ctrl} \gets S_{ctrl} + p$\;
            }
            \Else{
                $S_{ctrl} \gets S_{ctrl} - 2p$\;
                index $\gets 2$\;
            }
        }
        \Case{2}{
            \If{$I_\text{actual} > I_\text{prev}$}{
                $S_{ctrl} \gets S_{ctrl} - p$\;
            }
            \Else{
                $S_{ctrl} \gets S_{ctrl} + p$\;
                index $\gets 0$\;
            }
        }
    }
}
\end{algorithm}
\end{minipage}
\hfill
\begin{minipage}[t]{0.48\textwidth}
\begin{algorithm}[H]
\caption{Adaptive Perturb-and-Observe (P\&O) with Circular Constraint. Inputs:Current intensity $I_\text{actual}$, previous intensity $I_\text{prev}$, estimated maximum intensity $I_\text{max}$, adaptive parameters $\alpha, \beta$, enable signal $EnControl$, synchronization pulse $sync$, bounds $S_{\min}, S_{\max}$. Output: Control signal $S_{ctrl}$}
\label{alg:adaptive_po_circular}
\SetAlgoLined
\If{rising edge of $sync$}{
    $I_\text{prev} \gets I_\text{actual}$\;
    $dp = \frac{1}{8}(I_\text{max} - I(t)) + 50$\;
    \Switch{index}{
        \Case{0}{
            \If{$EnControl = 1$}{
                $S_{ctrl} \gets S_{ctrl} + dp$\;
                index $\gets 1$\;
            }
        }
        \Case{1}{
            \If{$I_\text{actual} > I_\text{prev}$}{
                $S_{ctrl} \gets S_{ctrl} + dp$\;
            }
            \Else{
                $S_{ctrl} \gets S_{ctrl} - 2dp$\;
                index $\gets 2$\;
            }
        }
        \Case{2}{
            \If{$I_\text{actual} > I_\text{prev}$}{
                $S_{ctrl} \gets S_{ctrl} - dp$\;
            }
            \Else{
                $S_{ctrl} \gets S_{ctrl} + dp$\;
                index $\gets 0$\;
            }
        }
    }
    \If{$S_{ctrl} > S_{\max}$}{
        $S_{ctrl} \gets S_{\min}$\;
    }
    \If{$S_{ctrl} < S_{\min}$}{
        $S_{ctrl} \gets S_{\max}$\;
    }
}
\end{algorithm}
\end{minipage}
} 
\caption{ Classical and Adaptive versions of the P\&O control algorithm.}\label{Fig_Algoritmos}
\end{figure}

In this scheme, the step size is modulated based on the distance between the current intensity $I(t)$ and the maximum intensity $I_\text{max}$. Specifically, the step is computed as: 
\begin{equation}
   dp =\frac{\alpha}{I_\text{max}} \left(I_\text{max} - I(t) \right) + \beta,
\end{equation}
where $\alpha$ sets the upper limit for the perturbation size, and $\beta$ defines the minimum allowable step size. For example, considering a 12-bit resolution, that is $4096$ quantization levels called Arbitrary Unit (AU), with step sizes ranging from a minimum of 50AU to a maximum of 562AU, the step update is given by 
$dp =\frac{1}{8} \left(I_\text{max} - I(t) \right) + 50$. 
Note that divisions in the FPGA are implemented in base 2 for simplicity, and the results are restricted to integer values.

The value of $I_\text{max}$ can be defined by the user or estimated dynamically during system operation. When the system operates far from the intensity maximum, the algorithm increases the step size to accelerate convergence. As the system approaches the optimum, the step size naturally decreases, reducing overshoot and improving stability. Additionally, the algorithm continuously evaluates the gradient of the intensity response in real time, enabling it to effectively navigate toward the optimal operating point.

As mentioned above, the controller has an operating range, which we can call a practical range constraint. This constraint further complicates the nonlinearity of the system. Therefore, in practical implementations, the control signal $S_{ctrl}$ is limited by the dynamic range of the actuator, typically between $S_{\min}$ and $S_{\max}$. To avoid saturation or blocking at the control limits, we have introduced a circular constraint that allows the signal to adjust to the operating range, maintaining continuous operation. Such a constraint is physically consistent with the $2\pi$ periodicity of the optical phase, allowing for uninterrupted stabilization of the interferometer.

Both algorithms rely exclusively on intensity feedback and are model-free, which simplifies real-time implementation in digital hardware platforms such as FPGAs.

\section{Results}

\subsection{Self-adjust the circular constraint}

An electrical sawtooth signal is employed to self-adjust the circular constraint. The lower limit is typically set to a value close to zero and can be defined by a User.
\begin{figure}[h]
\centering\includegraphics[width=11cm]{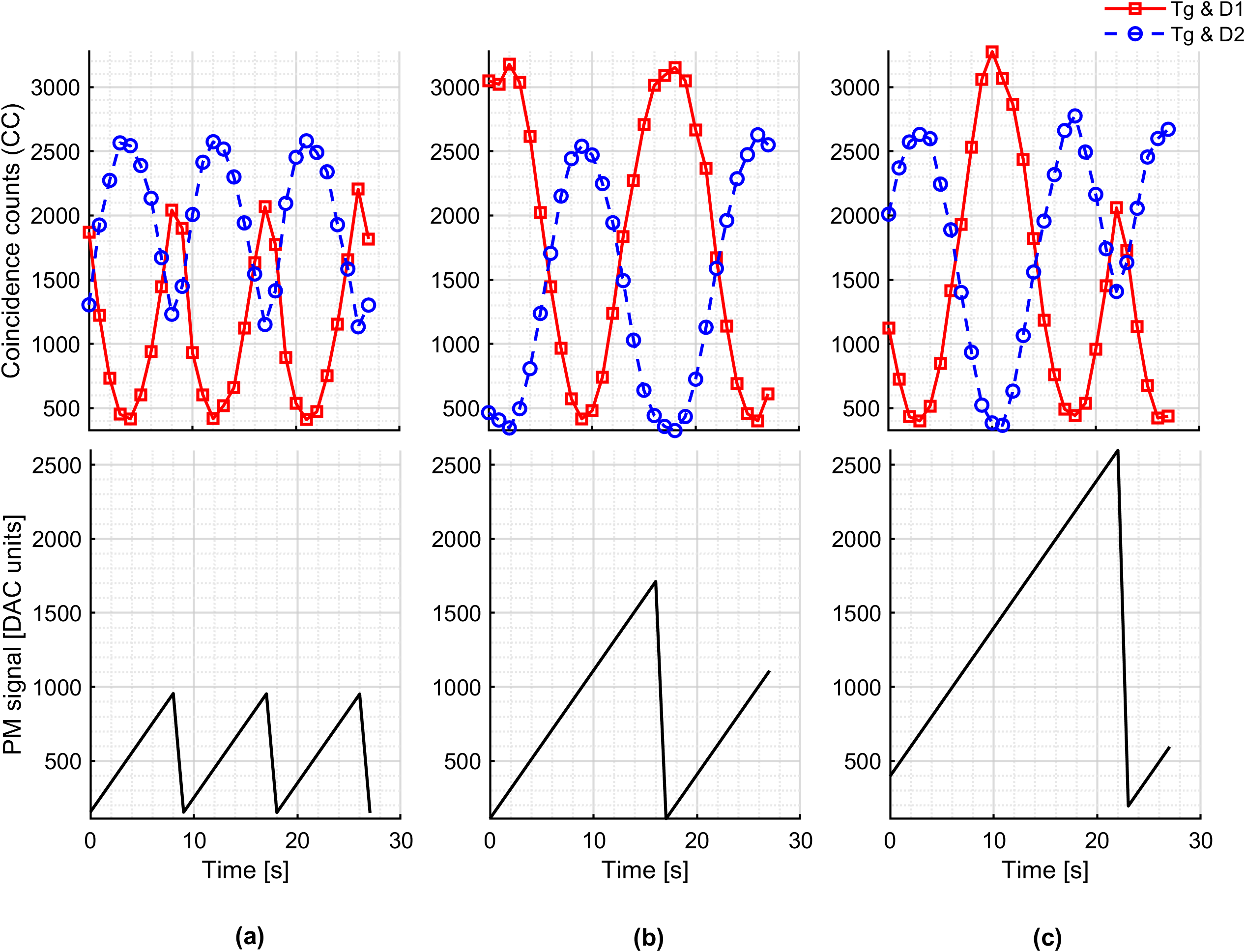}
\caption{Sawtooth signal used to self-adjust the circular constraint. Small amplitude yields incomplete modulation (a). Optimal amplitude ensures continuous cosine–sinusoidal output (b). Large amplitude causes discontinuities (c).}
\label{figalineacion}
\end{figure}
As shown in Figure~\ref{figalineacion}.a, when the sawtooth amplitude is too large, the optical output exhibits a discontinuity (exceeding one wavelength). Conversely, if the amplitude is too small, the signal does not achieve a full wavelength, leading to incomplete modulation (see Figure~\ref{figalineacion}.c ). By adjusting the amplitude, an optimal point of continuity is achieved in the output, following the expected cosine-sinusoidal trend (see Figure~\ref{figalineacion}.b). This procedure enables the system to self-adjust and establish the appropriate limits ($S_{\text{min}}$ and $S_{\text{max}}$), thereby ensuring uninterrupted operation.

\subsection{FPGA Performance}
The implementation was performed on a Xilinx Zynq-7000 FPGA device. 
Details of the utilization of FPGA resources for the adaptive P\&O and conventional P\&O approaches are provided in Table~\ref{table_fpgaComparison}. 
The results indicate that the adaptive algorithm, combined with circular constraint search, does not lead to a significant increase in the use of FPGA resources. 
Specifically, the adaptive version requires 3010 Look-Up Tables (LUTs) compared to 2949 in the conventional P\&O implementation, and 3556 Flip-Flops (FFs) compared to 3499. These increments correspond only to +0. 11\% and +0. 05\% of the total available device resources, respectively. For all other resources, including LUTRAM, Block RAM (BRAM), Digital Signal Processors (DSPs), Input/Output blocks (I/O), and Mixed-Mode Clock Managers (MMCMs), the utilization remains unchanged.
Thus, the adaptive algorithm can be incorporated without a meaningful impact on the overall FPGA resource budget, while still providing the benefits of adaptive control.
\begin{table}[h!]
\centering
\caption{FPGA resource utilization: Adaptive Algorithm vs. P\&O (percentage of available resources)}
\label{table_fpgaComparison}
\begin{tabular}{|l|c|c|c|c|}
\hline
\textbf{Resource} & \textbf{Available} & \textbf{Util. P\&O  } & \textbf{Util. Adaptive P\&O } & \textbf{Diff. (\% of Available)} \\ \hline
LUT      & 53200  & 2949  & 3010  & +0.11\% \\ \hline
LUTRAM   & 17400  & 1040  & 1040  & 0.00\%  \\ \hline
FF       & 106400 & 3499  & 3556  & +0.05\% \\ \hline
BRAM     & 140    & 0.50  & 0.50  & 0.00\%  \\ \hline
DSP      & 220    & 1     & 1     & 0.00\%  \\ \hline
IO       & 200    & 25    & 25    & 0.00\%  \\ \hline
MMCM     & 4      & 1     & 1     & 0.00\%  \\ \hline
\end{tabular}
\end{table}

\subsection{Experimental Phase Stabilization}

\begin{figure}[htbp]
\centering\includegraphics[width=13cm]{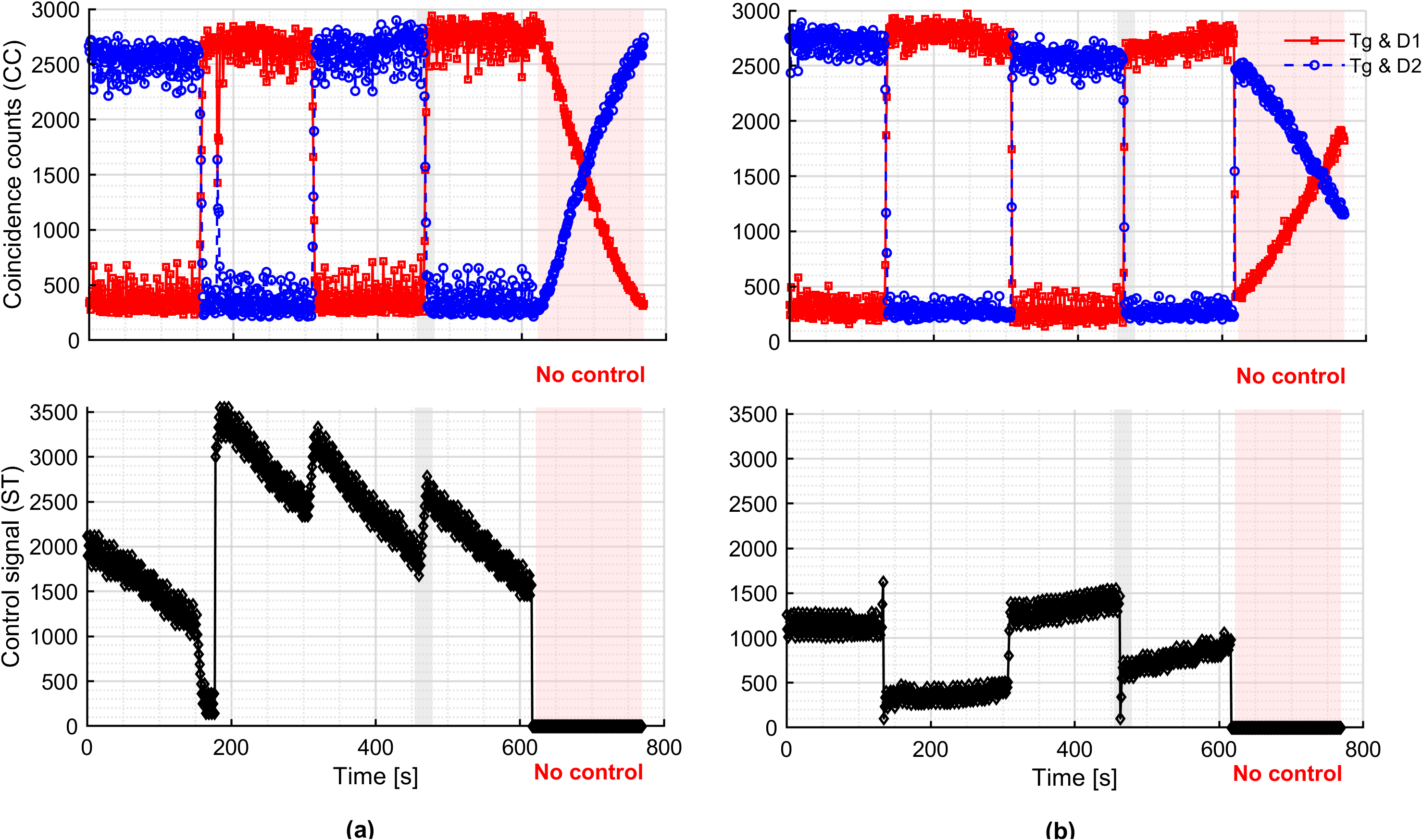}
\caption{Operation of the control system showing coincidence counts for controlled and uncontrolled intervals (top), and the performance of the conventional (a) and adaptive (b) P\&O algorithms at the stretcher output (bottom).}\label{Fig_sistemaControlado}
\end{figure}

Figure \ref{Fig_sistemaControlado} shows the operation of the control system. The upper graphs show the coincidence counts for three cases: (i) the system operating under controlled phase noise, (ii) a reference channel transition performed from Tg–D1 to Tg–D2, and (iii) a segment corresponding to the uncontrolled system (aprox. after 600 s). The latter region is highlighted with a red background to clearly indicate the interval where the system operates without phase control.
The lower graphs depict the operation of each control algorithm applied to the stretcher output, using 12-bit digital-to-analog converter (DAC) units. Figure \ref{Fig_sistemaControlado}(a) shows the system employing the conventional P\&O algorithm, whereas Figure \ref{Fig_sistemaControlado}(b) presents the system using the adaptive P\&O algorithm. Note that the conventional P\&O controller exhibits fixed step sizes, while the adaptive controller can vary its step size and remains immune to constraint-induced jumps when operating with the circular constraint, which, as shown in Figure \ref{figalineacion} (b), has an upper limit of 1700 ADU.

\subsection{Dynamic and Noise Behavior Comparison}
\begin{figure}[htbp]
\centering\includegraphics[width=11cm]{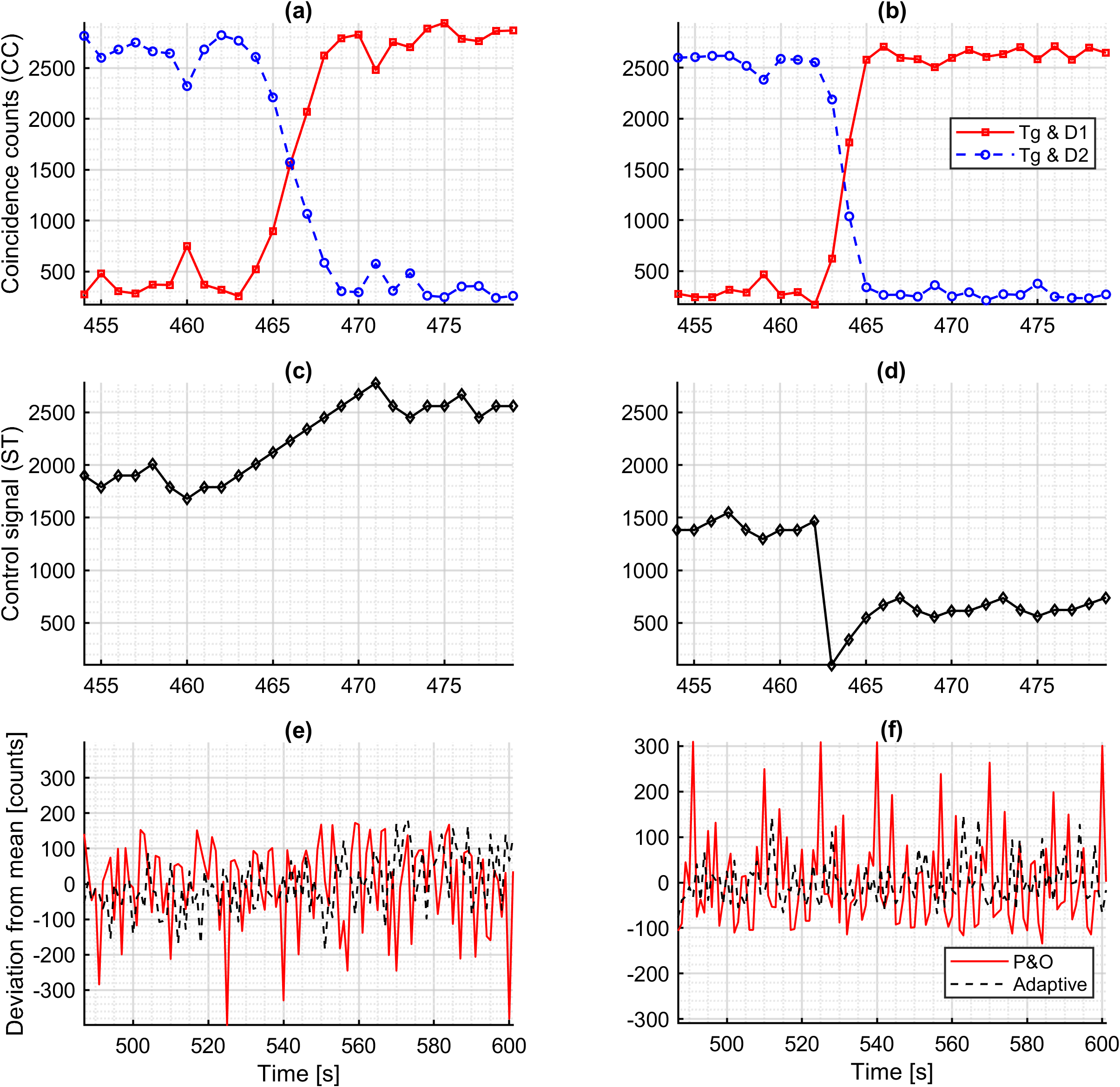}
\caption{Adaptive algorithm performance. (a–b) Control channel transition ($t = 85$–$97$~s) for conventional and adaptive P\&O algorithms. (c–d) 12-bit stretcher output showing fixed vs. adaptive step behavior. (e–f) Noise in the steady-state region ($t = 100$–$200$~s) for Tg–D1 and Tg–D2 channels.}
\label{fig_zoomRuido}
\end{figure}

To provide an example of the adaptive algorithm’s speed and stability behavior, two time segments were extracted from Figure \ref{Fig_sistemaControlado}. The first segment, between t = 454s and t = 479s, corresponds to the reference transition of the control channel. This interval is shown in Figures \ref{fig_zoomRuido} (a) and (b) for the conventional and adaptive P\&O algorithms respectively, where its rise and fall times can be clearly observed. The lower plots (Figures \ref{fig_zoomRuido}(c) and (d)) show the 12-bit output of the stretcher. These plots illustrate the fixed step size of the conventional controller and the adaptive behavior of the proposed algorithm, highlighting the nonlinear response of the adaptive system, which results in a reduced rise time compared to the conventional controller (see Figures \ref{fig_zoomRuido}(a) and (b)).
In contrast, Figures \ref{fig_zoomRuido} (e) and (f) show an enlarged view of the stable region between t = 487 s and t = 601 s, where the noise is evaluated with respect to the mean value obtained during stabilization. A maximum of 3000 coincidence counts (CC) was used as a reference level, intentionally set because the measured counts did not reach this stabilization threshold. Figure \ref{fig_zoomRuido} (e) shows the noise for the Tg–D1 channel, while Figure \ref{fig_zoomRuido} (f) corresponds to the Tg–D2 channel.

\textbf{The rise and fall times ($t_{\mathrm{rise-fall}}$)}  were obtained by analyzing the transient response of each control algorithm within specific time intervals of the experiment. Three representative transient windows were selected from Figure~\ref{Fig_sistemaControlado}, corresponding to [129-170] s, [300-320] s, and [454-479] s. The last interval is illustrated in Figures~\ref{fig_zoomRuido}(a) and~(b). Within each interval, the instants $t_{10}$ and $t_{90}$ denote the time coordinates at which the output signal reaches 10\% and 90\% of its final steady-state value, respectively. These time coordinates were used to calculate the rise or fall time as $t_{\text{rise/fall}} = |t_{90} - t_{10}|$. The results obtained from the intervals were then averaged to characterize the overall dynamic behavior of the proposed control.  
Another performance indicator that quantifies the relative improvement in rise time achieved by the adaptive P\&O algorithm compared to the conventional one is the percentage decrease, defined as:
\begin{equation}
 \Delta_{t_{r/f}} (\%) = 100 \times \frac{t_{\mathrm{rise,P\&O}} - t_{\mathrm{rise,Adaptive}}}{t_{\mathrm{rise,P\&O}}}
 \end{equation}
where $\Delta_{t_rise}(\%)$ represents the relative decrease in rise time achieved by the adaptive P\&O algorithm with respect to the conventional method. A positive $\Delta_{t_{r/f}}$ (\%) denotes reduced rise / fall times and increased operating speed.
According to Table~\ref{table_riseTime}, the Adaptive P\&O algorithm achieves an average reduction of approximately 60–70\% in the rise-fall time compared to the conventional P\&O method, indicating faster control loop stabilization and a more efficient adaptation to optical perturbations.

 \begin{table}[htbp]
\centering
\setlength{\tabcolsep}{6pt}
\renewcommand{\arraystretch}{1.1}
\caption{Average rise and fall time characteristics (10--90\%) of the P\&O and Adaptive P\&O algorithms obtained from the data shown in Figure~\ref{Fig_sistemaControlado}. The parameter $ \Delta_{t_{r/f}}$ (\%) represents the relative improvement in transient response achieved by the Adaptive P\&O algorithm.}
\begin{tabular}{@{}lccc@{}}
 \toprule
\textbf{Channel} & \textbf{P\&O $t_{\text{r/f}}$ [s]} & \textbf{Adaptive $t_{\text{r/f}}$ [s]} & \textbf{$\Delta_{t_{r/f}}$ (\%)} \\
\midrule
Tg \& D1 & $6.11 \pm 2.35$ & $3.18 \pm 2.01$ & $56.6 \pm 28.1$ \\
Tg \& D2 & $6.72 \pm 2.64$ & $2.61 \pm 0.66$ & $70.7 \pm 17.3$ \\
\bottomrule
\end{tabular}
\label{table_riseTime}
\end{table}

\textbf{Noise characterization:} signal stability was evaluated under steady state conditions in four time intervals: [1--130]~s, [169--243]~s, [339--446]~s, and [487--601]~s (see Figures~\ref{fig_zoomRuido}(e) and~(f)). 
Within each interval, the statistical fluctuations of the coincidence counts were characterized by the mean rate ($\mu$) and standard deviation ($\sigma$), extracted directly from the experimental data. 
From these quantities, the expected Poisson noise was calculated as $\sigma_\mathrm{P} = \sqrt{\mu}$, and the coefficient of variation as $\mathrm{CV} = \sigma / \mu$, representing the normalized noise amplitude. 
The Poisson noise served as a theoretical reference in photon-counting statistics. 
In order to compare the noise performance of both control methods, we employ the percentage decrease, defined as:
\begin{equation}
\Delta_{\sigma}(\%) = 100 \times \frac{\sigma_\mathrm{P\&O} - \sigma_\mathrm{Adaptive}}{\sigma_\mathrm{P\&O}},
\label{eq_IRN}
\end{equation}
where $\sigma_\mathrm{P\&O}$ and $\sigma_\mathrm{Adaptive}$ denote the standard deviation values obtained from the classical and adaptive P\&O methods, respectively.
A positive $\Delta_{\sigma}(\%)$ denotes reduced noise, similar to the time indicator.

This metric quantifies the relative reduction in output noise or oscillations achieved by the adaptive controller compared to the conventional approach.
The results obtained for each operating interval are summarized in Table~\ref{table_ruido}.

\begin{table}[htbp]
\centering
\setlength{\tabcolsep}{5pt}
\renewcommand{\arraystretch}{1.1}
\caption{Noise analysis of the P\&O and Adaptive P\&O algorithms across four operating intervals. Reported parameters include mean signal, standard deviation (Std), estimated Poisson noise ($\sigma_P$), coefficient of variation (CV), and Noise Improvement Ratio (IRN).}
\begin{tabular}{@{}lcccccc@{}}
\toprule
\textbf{Channel} & \textbf{Algorithm} & \textbf{Mean} & \textbf{Std} & \textbf{$\sigma_P$} & \textbf{CV} & \textbf{$\Delta_{\sigma}(\%)$} \\
\midrule
\multicolumn{7}{l}{\textbf{Noise analysis in [1--130] s}} \\
Tg \& D1 & P\&O     & 366.777 & 107.335 & 19.151 & 0.293 & -- \\
Tg \& D1 & Adaptive & 300.468 &  77.065 & 17.334 & 0.256 & +28.20 \\
Tg \& D2 & P\&O     & 2543.892 & 116.536 & 50.437 & 0.046 & -- \\
Tg \& D2 & Adaptive & 2708.715 &  84.920 & 52.045 & 0.031 & +27.13 \\
\midrule
\multicolumn{7}{l}{\textbf{Noise analysis in [169--243] s}} \\
Tg \& D1 & P\&O     & 2645.080 & 227.227 & 51.430 & 0.086 & -- \\
Tg \& D1 & Adaptive & 2796.942 &  65.155 & 52.886 & 0.023 & +71.33 \\
Tg \& D2 & P\&O     & 393.493  & 223.733 & 19.837 & 0.569 & -- \\
Tg \& D2 & Adaptive & 274.613  &  46.523 & 16.571 & 0.169 & +79.21 \\
\midrule
\multicolumn{7}{l}{\textbf{Noise analysis in [339--446] s}} \\
Tg \& D1 & P\&O     & 363.296 & 115.472 & 19.060 & 0.318 & -- \\
Tg \& D1 & Adaptive & 264.788 &  76.730 & 16.272 & 0.290 & +33.55 \\
Tg \& D2 & P\&O     & 2636.713 & 132.351 & 51.349 & 0.050 & -- \\
Tg \& D2 & Adaptive & 2566.083 &  80.314 & 50.657 & 0.031 & +39.32 \\
\midrule
\multicolumn{7}{l}{\textbf{Noise analysis in [487--601] s}} \\
Tg \& D1 & P\&O     & 2763.383 & 124.419 & 52.568 & 0.045 & -- \\
Tg \& D1 & Adaptive & 2696.272 &  79.334 & 51.926 & 0.029 & +36.24 \\
Tg \& D2 & P\&O     & 344.339 & 106.424 & 18.556 & 0.309 & -- \\
Tg \& D2 & Adaptive & 268.783 &  47.935 & 16.395 & 0.178 & +54.96 \\
\bottomrule
\end{tabular}
\label{table_ruido}
\end{table}

\section{Performance Improvement Analysis}

The results summarized in Tables~\ref{table_riseTime}--\ref{table_ruido} demonstrate that the Adaptive P\&O algorithm significantly improves both the dynamic and noise performance of the optical control system. 
The average rise--fall time decreases by approximately 57\% for Tg~\&~D1 and 71\% for Tg~\&~D2, indicating faster transient response and enhanced loop stabilization compared to the conventional P\&O scheme.
Under steady-state operating conditions, the Adaptive algorithm exhibits a pronounced reduction in statistical fluctuations across all analyzed intervals.
The coefficient of variation decreases from $0.186 \pm 0.125$ to $0.150 \pm 0.124$ for Tg~\&~D1, and from $0.244 \pm 0.242$ to $0.102 \pm 0.070$ for Tg~\&~D2.
This improvement is consistent with the corresponding percentage decrease of $42.8 \pm 19.3\%$ and $50.2 \pm 22.1\%$ for Tg~\&~D1 and Tg~\&~D2, respectively, confirming the enhanced signal stability provided by the adaptive control scheme.
In addition, the optical visibility of the system, determined from the difference between the maximum and minimum interferometric intensities, improved by approximately 7. 15\% when the adaptive P\&O algorithm was used compared to the conventional one.
These results confirm that the adaptive feedback scheme enhances interference contrast and optical phase stability while effectively suppressing the noise components.

\section{Conclusion}

We have experimentally demonstrated an adaptive Perturb-and-Observe (P\&O) phase stabilization scheme operating solely on coincidence counts in a time-bin quantum states system. 
The adaptive control achieved up to a 70\% reduction in transient response time and more than 50\% average noise suppression compared to the conventional P\&O algorithm, evidencing faster convergence and improved stability in steady state. 
These results confirm that coincidence-based adaptive control can effectively mitigate phase noise and statistical fluctuations. 
The proposed method is fully autonomous, reference-free, and readily scalable, offering a robust and resource-efficient approach for phase stabilization in large-scale and field-deployable quantum communication systems.

\begin{backmatter}
\bmsection{Funding}Fondo Nacional de Desarrollo Científico y Tecnológico (ANID) (Grants No. 1240843,  1240746); Ingeniería 2030 (ING222010004); Conselho Nacional de Pesquisa e Desenvolvimento Científico e Tecnológico (CNPq - process no. 422300/2021-7); Fundação de Amparo à Pesquisa do Estado de Minas Gerais (FAPEMIG - project no. BPD-00996-22); Centro de Pesquisas Avançadas Wernher Von Braun.

\bmsection{Acknowledgment}The authors acknowledge helpful discussions with Guilherme B. Xavier during the preparation of this article.

\bmsection{Disclosures}The authors declare no conflicts of interest.

\bmsection{Data Availability Statement}  Data underlying the results presented in this article are not publicly available at this time, but may be obtained from the corresponding authors upon reasonable request.
\end{backmatter}



\end{document}